\documentclass[12pt]{article}

\usepackage{amsmath}
\usepackage{fullpage}
\usepackage{graphicx}
\newcommand*{\dW}{\ensuremath{d\mathcal W}}
\newcommand*{\openone}{\ensuremath{\mathbf 1}}

\newcommand*{\avg}[1]{\ensuremath{{\left\langle{#1}\right\rangle}}}

\newcommand*{\dA}[1]{\ensuremath{\tfrac{\delta\mathcal A}{\delta x({#1})}}}
\renewcommand*{\H}{\ensuremath{\mathcal H}}
\newcommand*{\eval}[2]{\ensuremath{\left.{#1}\right|_{#2}}}
\newcommand*{\pd}[2]{\ensuremath{\frac{\partial {#1}}{\partial {#2}}}}
\renewcommand*{\Pr}[1]{\ensuremath{\mathcal{P}\left({#1}\right)}}
\newcommand*{\Pri}[2]{\ensuremath{\mathcal{P}_{#1}\left({#2}\right)}}
\newcommand*{\citenum}[1]{\cite{#1}}

\author{ David M. Rogers and Susan B. Rempe}
\title{ A First and Second Law for Nonequilibrium Thermodynamics: Maximum Entropy Derivation of the Fluctuation-Dissipation Theorem and Entropy Production Functionals}

\begin{document}

\maketitle
\begin{abstract}
  We derive a physically motivated theory for non-equilibrium systems from a maximum entropy approach similar in spirit to the equilibrium theory given by Gibbs.  Requiring Hamilton's principle of stationary action to be satisfied on average during a trajectory, we derive constraints on the {\em transition} probability distribution which lead to a path probability of the Onsager-Machlup form.  Additional constraints derived from energy and momentum conservation laws then introduce heat exchange and external driving forces into the system, with Lagrange multipliers related to the temperature and pressure of an external thermostatic system.  The result is a fully time-dependent, non-local description of a nonequilibrium ensemble coupled to reservoirs at arbitrary thermostatic states.  Detailed accounting of the energy exchange and the change in information entropy of the central system then provides a description of the entropy production which is not dependant on the specification or existence of a steady-state or on any definition of thermostatic variables for the central system.  These results are connected to the literature by showing a method for path re-weighting, creation of arbitrary fluctuation theorems, and by providing a simple derivation of Jarzynski relations referencing a steady-state.  In addition, we identify path free energy and entropy (caliber) functionals which generate a first law of nonequilibrium thermodynamics by relating changes in the driving forces to changes in path averages.  Analogous to the Gibbs relations, the variations in the path averages yield fluctuation-dissipation theorems.  The thermodynamic entropy production can also be stated in terms of the caliber functional, resulting in a simple proof of our microscopic form for the Clausius statement.  We find that the maximum entropy route provides a clear derivation of the path free energy functional, path-integral, Langevin, Brownian, and Fokker-Planck statements of nonequilibrium processes.  Physical considerations justify a fundamental definition of thermodynamic entropy increase as system information entropy plus heat exchange with an external thermostatic system.
\end{abstract}

  The Green-Kubo fluctuation theorems\cite{rkubo66} relate equilibrium time-correlation functions with the time-response of system observables to an external driving force.
They are important for their ability to calculate the transport coefficients appearing in Onsager's phenomenological relation~\cite{onsager31},
\begin{equation}
\label{eq:onsager}
\avg{\dot a_i(t)} \approx \sum_{ij} L_{ij} X_j(t)
,
\end{equation}
which identified $X_j$ with the ``entropic" driving forces, $\partial S / \partial a_j \approx \beta(t)-\beta_\text{eq}$\cite{callen}.

  These two relations form a rough draft for the a first law of nonequilibrium thermodynamics.  Combined with the usual second law prescription of increasing entropy, the above establishes the direction in which the system will relax toward equilibrium and an estimate of the entropy increase by this process.  However, the linear transport equations have been derived by analogy with the equilibrium theory, and their interpretation must be made with respect to the entropy of a quasi-steady process.  A more satisfactory development would therefore determine the range of conditions for which Eq.~\ref{eq:onsager} holds, as well as provide a foundation for studying processes with arbitrary driving forces and defined without reference to any equilibrium state.  If, in addition, the theory was able to resemble the well-known equilibrium statistical mechanics, it would offer a wealth of immediate insight into new applications for which now standard nonequilibrium methods may prove cumbersome and error prone.  Such a resemblance must be reached by defining path functionals analogous to the energy and entropy of equilibrium states -- and would therefore constitute a true statistical mechanics for thermodynamics (as opposed to thermostatics).  Our first question in this investigation will be how such a microscopic first law of thermodynamics can be formulated.

  Although much work has been devoted to the above problems, these questions have been addressed from a large number of different viewpoints in the last thirty years.  Further, there appears to be little consensus on a unifying, general, set of relations from which which all nonequilibrium results may be derived.\cite{etrep04}
  Recent work has centered on a second question of deriving a microscopic second law of thermodynamics through proving the existence of fluctuation theorems\cite{gcroo00} and exploring their consequences for nonequilibrium systems.

  A fundamental fluctuation theorem result is an expression for the ``lost work'' over and above the equilibrium free energy for stochastic processes that convert one thermostatic state to another\cite{cjarz97} via mechanical driving.  This lost work can be interpreted as an entropy increase.  It is simple to show in the case of an isolated deterministic system\cite{cjarz00}
\begin{equation}
\label{eq:jarz}
 e^{\beta_0(W(\{x\}_0^S)-\Delta F)} = \Pri{0}{x_0}/\Pri{S}{x_S(x_0)}
\end{equation}
Where the work, $W$, is the gain in system energy $\Delta U=U_S(x_S)-U_0(x_0)$, and both distributions are required to be at the same temperature, $k_B/\beta_0$, to use the equilibrium relation $\beta_0 \Delta F = -\ln Z_S/Z_0$.\cite{bpalm07}
Further work has provided examples of many fluctuation theorems.\cite{etrep04,vcher06}
Each can be used to define a measure of irreversibility, since for any two path probabilities, $A$ and $B$,
\begin{align*}
 e^{l(\{x\}_0^S)} &\equiv \Pri{B}{\{x\}_0^S}/\Pri{A}{\{x\}_0^S} \\
 \Rightarrow \avg{l}_B &= \mathcal D[B|A] \ge 0
,
\end{align*}
where $\mathcal D$ is the Kullback-Leibler divergence (and necessarily positive).  Although special significance is often attributed to time reversal (defined by replacing the time evolution operator from $i$ to $i+1$ with a reversal of odd functions of time at step $i+1$, unaltered time evolution from $x^*_{i+1}$ to $x^*_i$, and another reversal of odd functions at $i$),\cite{cmaes99} this operation is not possible if one-way steps are present, so that only one of $i\to i+1$ or $i+1^*\to i^*$ has probability zero.  Furthermore, the extension of this formalism to systems lacking momentum, such as discrete processes described by a transition probability matrix is also unclear (although suggestions involving a [possibly non-unique] stationary distribution) have been offered\cite{gcroo00}.  Until these issues are resolved, there does not exist a definitive path functional able to answer the first question above.



  Early works were principally focused on the first question.  After a detailed picture of how time-correlations control the rate of relaxation to equilibrium\cite{jkirk46,hcall51}, a general set of relations (projector-operator theory\cite{rzwan61,rzwan72,hmori65,rzwan65}) was described from which all such macroscopic relaxations may be derived.
The principal content of the theory was to define a coarsening, projection, operator which removes information about the unmodeled degrees of freedom.  The exact, non-Markovian kinetic equation for the probability distribution in the space of remaining variables then shows the relaxation process in the form of a time convolution of the time-correlation functions and the thermodynamic forces driving them to equilibrium.\cite{rzwan61,rzwan72}  That is, Eq.~\ref{eq:onsager} should be replaced with
\begin{equation}
\label{eq:zwanzig}
\avg{\dot a_i(t)} \approx -\int_0^t \sum_{ij} L_{ij}(t-\tau) X_j(t-\tau) d\tau
,
\end{equation}
where $L_{ij}(s) = \avg{\dot a_i(t) \dot a_j(s)}$.
Despite conjectured relationships to a nonequilibrium entropy function,\cite{lonsa31,lonsa53,rzwan65} its relation to the thermostatic entropy has not yet been fully justified in terms of the maximum entropy formalism or clarified to the point where it is possible to derive the above from a second derivative\cite{ejayn80} -- analogous to the Gibbs relations for equilibrium which give rise to obviously related quantities such as heat capacity, compressibility and coefficient of thermal expansion.


  At this point, the situation shared a peculiar similarity to the circumstances surrounding Gibb's classical text introducing the principle of maximum entropy\cite{jgibb02}.  The method for deriving fluctuation dissipation theorems (analogous to the first law) was to define a system evolving according to an exact Lagrangian, make a random phase approximation to yield an ensemble of exactly evolving trajectories, and then derive a corresponding ``physical'' distribution on trajectory space (analogous to phase-space, Tbl.~\ref{tbl:corresp}).  Because of the prevailing attitude regarding mechanics as the only possible method for solving such problems at the time, Gibbs use of maximum entropy methods was seen as a non-physical trick\cite{eschr67} to derive properties of molecular equilibrium.  Again, introducing maximum entropy lead to an expansive generalization of the fluctuation dissipation theorem by Jaynes'\cite{ejayn80,droge11a} -- who, it should be noted, has recognized and written about the above conflict of Gibbs\cite{ejayn79}.

  Jaynes introduced maximum entropy following the program of Gibbs and using the subjective interpretation of probability used by Laplace and Jeffreys.\cite{ejayn03}  Making the substitution from a state (containing a variable at a single time) to an entire trajectory immediately identifies a non-equilibrium analog of the Gibbs ensemble, complete with entropy and free energy functionals on trajectory space.  Pursuing the analogy further, Jaynes showed that the first derivatives of the path free energy yielded averages of path functionals, and the second derivatives (minus the space and time-correlation functions) give their ``first-order'' perturbation with respect to changing the thermodynamic forces, $\lambda$\cite{ejayn79,ejayn80,ejayn85}.  Therefore we can shorten Eq.~\ref{eq:zwanzig} to
\begin{equation}
\label{eq:jaynes}
 \avg{a} \approx \avg{a}_0 - \sum_k \avg{(a-\avg{a}_0) (b_k-\avg{b_k}_0)}_0 (\lambda_k-\lambda^0_k)
,
\end{equation}
for {\em any} path functional, $a$, we wish to average and {\em any} set of ``controlled'' quantities $b_k$ which define our trajectory ensemble.  If, for example, $b_k$ included time and space-dependent particle fluxes (i.e. $k$ indexes both time and space so that $\sum_k\to\iint dt dx$), then Eq.~\ref{eq:jaynes} easily generalizes Eq.~\ref{eq:onsager} to fluctuation relations defined directly from the set of constrained path functionals, $\avg{b_k}$.  Notice that the perturbation expansion above is not defined by reference to an equilibrium state, but instead with respect to a reference probability distribution on path space.  Examples of such elegant derivations of fluctuation theorems have been given many times in works employing path integral methods\cite{lchen04} such as those related to the Onsager-Machlup action\cite{lonsa53,khunt81}.  Further work\cite{cmaes99,jlebo99,rdewa03} has also shown how Jaynes' path entropy may be connected to fluctuation and entropy production theorems.

  However, Eq.~\ref{eq:jaynes} contains a fatal flaw.  To see this, first note that the average of a quantity at time $t$ is dependent on the ``control'' parameters throughout the whole trajectory.  This is because logical inference does not contain a preferential time direction, and knowledge that the system has a given property at time $t+\tau$ constrains the state at time $t$.  Although this could be alleviated by requiring only casual information to enter into the determination of the state at time $t$, this approach leads to a probability distribution valid only for $x_t$ and not any previous times.  A better approach is to maximize the entropy of the transition probability distribution.

  Limiting the scope of the maximum trajectory entropy procedure in this way automatically corrects a related re-normalization problem.  Suppose phase space were to branch at a future time $t+\tau$.  In this case, a uniform measure on path space would assign points at time $t$ a different weight depending on future events.  A simple example is the Monty Hall problem with a prize assigned to the first door without loss of generality.\cite{jgill02}  The contestant's choice of one of three doors plus Monty opening another (not concealing a prize) constitutes four possible paths, and maximum path entropy would give each path an equal weight -- an intuitive, but incorrect, solution.  The correct probability assignment is a uniform distribution for each {\em transition}, leading to ending weights of $1/3 \times 1/2=1/6$ for the two paths following the selection of the first door\footnote{We leave updating the player's state of knowledge about the prize out of the discussion, and reiterate that the full path probabilities are $1/6,1/6$, (contestant's correct choice plus Monty choosing randomly from the remaining doors), and $1/3,1/3$ (contestant's two incorrect choices plus Monty's forced choice of an incorrect door).}.  In more abstract terms, the marginal probability at each time should not depend on the future -- a concept expressed mathematically by defining a progressively measurable function with respect to the natural filtration of stochastic processes\cite{abobr05}.

  In practice, this flaw can be avoided by considering only processes where assigning equal {\em a priori} weight to all paths is equivalent to assigning equal weight to all transitions.  By Liouville's theorem, this is obviously true for deterministic processes.  More generally, the approaches are equivalent when the number of possible transitions does not depend on the starting point.  The path entropy approach of Jaynes is therefore valid in the absence of factors re-normalizing for starting-point dependent differences in the number of possible paths ($Z[\lambda_i,x_i] = Z[\lambda_i]$ in Eq.~\ref{eq:Z}).

  A further question remains on the application of fluctuation-dissipation theorems to both Langevin and Brownian (overdamped Langevin) processes.  What is the appropriate order for coarse-grained equations of motion?  Writing down a Langevin equation assumes Newtonian, second-order dynamics, whereas the first-order Brownian motion can also be derived via the same approach.  Although the Green-Kubo relations were supposed to have solved these problems, ambiguity remains from this approach at a fundamental level because a strict derivation using the method of mechanics does not offer direct insight into the choice of macroscopic variables used for systematic coarse-graining.  In fluid mechanics, either can be applied, and the choice between full Newtonian motion models or simplified advection-diffusion models is predictably based on the scales of length, relaxation time, and applied force involved.
Several authors have made substantial contributions\cite{jboon91,rzwan01,gkarn05,devan08} in connecting fluid mechanics and non-equilibrium dynamics.

  Although seemingly incompatible, there are advantages to both the mechanics-based and the statistical derivations.  For example, considering the choice of equations of motion from the statistical perspective of Equation~\ref{eq:jaynes} at once allows us to see the consequences of each choice.  If we choose a first-order equation of motion, then we substitute the velocity at a point for $a$ and expand it about a streaming velocity $\avg{a}_0$ in the particle fluxes, $b_j(x,t)$.  If second order, then $a$ becomes a change in momentum at a point and we expand about the average force, $\avg{a}_0$, in the stress tensor of the surrounding fluid, $b(x,t)$.
We believe that further corrections sometimes employed in fluid mechanics may also be derived via extending the process that lead to Eq.~\ref{eq:jaynes}.  If both the nonequilibrium first and second laws could be addressed from the same perspective, it seems that an expansive generalization of non-equilibrium statistical mechanics could be achieved.

  In order to combine these two viewpoints, we introduce the following maximum entropy argument.  Suppose the information about the time change of a set of dynamic variables, $x$, consists in a set of prescribed averages,
$\{\avg{f_k(x_{i+1},x_i) | x_i}\}_{k=1}^m$.  Using only this information, we are to construct a probability distribution for the dynamic variables at point $i+1$, given a known $x$ at time $i$.  According to the standard maximum information entropy ($\H$) machinery, the answer is
\begin{align}
\begin{split}
\label{eq:rSmax}
\delta \H_{i+1|i} &= \delta \sum_{\{x_{i+1}\}} p(x_{i+1}|x_i) \Big[-\ln \frac{p(x_{i+1}|x_i)}{p^0(x_{i+1}|x_i)}\\
  & - (\ln Z[x_i] - 1) - \sum_{k=1}^m \lambda_{k,i} f_k(x_{i+1},x_i) \Big]
\end{split} \\
\label{eq:p}
p(x_{i+1}|x_i) &= p^0(x_{i+1}|x_i) e^{-\eta_i[x]} / Z[\lambda_i,x_i] \\
\label{eq:eta}
\eta_i[x] &\equiv \sum_{k=1}^m \lambda_{k,i} f_k[x_{i+1},x_i] \\
\label{eq:Z}
Z[\lambda,x_i] &= \sum_{\{x_{i+1}\}} p^0(x_{i+1}|x_i) e^{-\eta_i[x]} \\
\label{eq:rS}
\H_{i+1|i} &= \ln Z[\lambda,x_i] + \avg{\eta_i|x_i} \\
\label{eq:drS}
d\H_{i+1|i} &= \sum_{k=1}^m \lambda_k d\avg{f_k|x_i} \\
\label{eq:F}
\Psi_{i+1|i} &\equiv \sum_{k=1}^{m'} \lambda_k \avg{f_k|x_i} - \H_{i+1|i} \\
\label{eq:dF}
d\Psi_{i+1|i} &= \sum_{k=1}^{m'} \avg{f_k|x_i} d\lambda_j - \sum_{k=m'+1}^m \lambda_k d\avg{f_k}
. 
\end{align}
These are the expressions relating to the statistical state at time $i+1$ given information on the transition probability distribution.  The functional notation for quantities such as $\eta_i[x]$ has been used to indicate that in general, these may be considered as functionals depending on the trajectory over all times before $i+1$.  In the continuous limit, the above quantities exist between times $i$ and $i+1$, and should be viewed in the Stratonovich definition.  The appendix shows example calculations of the partition function for the Wiener process.

  The essential difference between Eq.~\ref{eq:p} and the maximum path entropy prescription is that in the latter, the path probability is
\begin{align*}
\Pr{\{x_i\}_0^S} &= \frac{e^{-A[\{x_i\}_0^S]-\beta U(x_0)}}{Z} \Pri{0}{\{x_i\}_0^S}
, \\
\intertext{for some path functional, $A$, and constant, $Z$, while this is replaced by}
\Pr{\{x_i\}_0^S} &= \prod_{i=0}^{S-1} \frac{e^{-\eta_i[x_{i+1};x_i]}}{Z_i[x_i]} \Pri{0}{\{x_i\}_0^S}
\end{align*}
in the former.  The present conditional expansion implies a dual characterization of a stochastic process as a single ensemble of paths and a set of telescoping ensembles, each with well-defined, non-anticipating energy changes through $\Delta F_k = \sum_{I=0}^{J-1} \avg{f_k[x_{i+1},x_i]}$.

\begin{table}[htbp]
 \centering
 \begin{tabular}{ll}
  Equilibrium & Non-Equilibrium \\
\hline
  Phase Space & Trajectory Space \\
  Free Energy & Path Free Energy \\
  Entropy & Caliber \\
  Average Value Constraint & Average Flux Constraint \\
  Equilibrium Average & Path Average \\
  Conditional Free Energy (PMF) & Conditional Path log-Probability \\
  Thermodynamic forces ($\delta$PMF) & Changes in Path Flux (work) \\
  Heat Capacity & Thermal Conductivity \\
  (none)       & Irreversibility and Entropy Production
 \end{tabular}
 \label{tbl:corresp}
 \caption{Correspondence between single-time and time-dependent path maximum entropy formulations of statistical mechanics.}
\end{table}

  Using the above, maximum transition entropy, form has several distinct advantages for
the derivation of non-equilibrium relations.  Not least is the correspondence to the canonical, maximum entropy form of equilibrium thermostatics pointed out in Table~\ref{tbl:corresp}.  We begin by deriving the
generalized Langevin and Brownian dynamics from a consideration of the action
deviation as a constrained quantity.
In Sec.~\ref{sec:thermo}, we connect the Langevin equations derived from our stochastic
action deviation principle to energy exchange and thermodynamic entropy production via
interaction with external reservoirs.\cite{pberg55}  Following Jaynes'
information theoretic derivation of statistical mechanics, we identify a quantity analogous
to a free energy and entropy functionals for maximum entropy transition probability processes in Sec.~\ref{sec:pred}.
The coefficient of thermal expansion, isothermal compressibility, and heat capacity can be derived using second derivatives of the equilibrium free energy.  In an analogous way, we show how Green-Kubo transport theory
can be derived from second derivatives of Legendre transforms of the maximum transition entropy functional.
These relations can describe the response of both steady and non-steady states arbitrarily far from
equilibrium.  Applications of this result enable calculation of derivatives of the current-voltage
curve at constant current or constant voltage and give general conditions under which Onsager
reciprocity will hold.
We then connect these functionals to the thermodynamic entropy production and Crooks-Jarzynski fluctuation relations.

\section{ Dynamic Constraints}
\label{sec:dyn}

  Starting from a mechanics problem specified by Lagrangian, $L(x,\dot x)$,
the usual mechanical prescription is to require stationary action.
\begin{align}
 \mathcal{A} &\equiv \int_0^S L(x,\dot x,t) dt \notag \\
 \delta\mathcal{A} &= \int_0^S \pd{L}{x} \delta x + \pd{L}{\dot x} \delta \dot
x \; dt + \left.L\right|_0^S \notag \\
  &= \left.\pd{L}{\dot x} \delta x\right|_0^S + \left.L\right|_0^S
    + \int_0^S \left(\pd{L}{x}-\frac{d}{dt} \pd{L}{\dot x} \right) \delta x(t) \;
dt \notag \\
\label{eq:dAdx}
  \Rightarrow& \frac{\delta\mathcal A}{\delta x(t)} = \pd{L}{x} - \frac{d}{dt} \pd{L}{\dot x}
 = F - \dot p
\end{align}
In this report, we will freely substitute force, $F\equiv\pd{L}{x}$, and momentum, $\dot p\equiv\frac{d}{dt} \pd{L}{\dot x}$.  The requirement for stationary action then reads $\dA{t}=^\text{set} 0 \Rightarrow F=\dot p$.

  When $L=\dot x^T M \dot x/2 - U(x)$, the above procedure directly gives
Newtonian mechanics and has the advantage of being generally
valid under coordinate transformations, $y = y(x), x = x(y)$.  However Eq.~\ref{eq:dAdx}
gives second-order equations of motion, requiring $\dot x = dx(t)/dt$ by
definition, and complicating discussions of numerical integration.  A first-order form can be derived from an alternate Lagrangian,
\begin{equation}
\label{eq:L1}
L(\begin{bmatrix}q\\v \end{bmatrix},\begin{bmatrix}\dot q\\ \dot v\end{bmatrix})
  = \left(\dot q - \tfrac{1}{2}v\right)^T M v - U(q)
,
\end{equation}
by treating $q(t)$ and $v(t)$ as separate dynamic variables.  Using
$x(t)=\{q(t),v(t)\}$ in (\ref{eq:dAdx}), we find
\begin{align}
\label{eq:dAdq}
\frac{\delta\mathcal{A}}{\delta q} &= -\pd{U}{q} - M\dot v \\
\label{eq:dAdv}
\frac{\delta\mathcal{A}}{\delta v} &= M(\dot q - v)
.
\end{align}
Setting these two equal to zero gives a result entirely equivalent to Newtonian
mechanics, but in which we may consider a Verlet-type integration process of
updating $v$ with fixed $q$, and then updating $q$ with $v$ fixed.
It has been found that Eq.~\ref{eq:dAdx} forms a
solid basis for forming generalizations of physical laws.

  A stochastic generalization may be to permit small deviations by constraining $f=\tfrac{\delta\mathcal{A}}{\delta x_i(t)}\tfrac{\delta\mathcal{A}}{\delta x_j(t-\Delta t)}\epsilon$ in (\ref{eq:p}).  The resulting matrices of constraint values (Lagrange multipliers $G_{\Delta t}, \Delta t \in [0,\infty)$) can restrict deviations in the action in a history-dependent and non-local way.  Each deviation in the action can be thought of as arising from elastic collisions with un-modeled molecules from the surrounding `bath' environment or as an unknown Lagrangian applied to the system between times $t-\epsilon$ and $t$\cite{ejayn57a} (with $\epsilon$ a small time increment).  Without any other constraints, this squared-deviation constraint implies interaction with a completely chaotic (infinite temperature) bath and does not conserve energy, momentum, etc. except in the deterministic limit ($\mathbf G \to I\infty$).

  When two systems are coupled, the combined system should obey a set of conservation laws.  As formalized by Noether's theorem,\cite{enoet71} such conservation laws can be derived for a single system directly from the action formulation by considering continuous transformations of the trajectory $x(t) \to x(t) + q(t,a)$ in a region around $a=0$ -- where $q(t,0) = 0$.  Because the action is stationary with respect to small perturbations in $x(t)$, there exists a vanishing quantity,
\begin{align}
 \frac{d I(t)}{dt} &\equiv -\eval{\pd{\mathcal A}{a}}{a=0} \notag \\
\label{eq:da}
  &= -\eval{\pd{q(t,a)^T}{a}}{a=0} \frac{\delta\mathcal A}{\delta x(t)}
,
\end{align}
(with $T$ reserved for denoting transposition) and it is possible to define an invariant (using $\eval{\pd{q(t,a)}{a}}{a=0} \equiv y(t)$).
\begin{equation*}
I(t) - I(\infty) = \int_{-\infty}^t -y^T(t') \dA{t'} dt'
\end{equation*}
Feynman showed that if the action functional is invariant to this transformation ($\mathcal A[x(t)] = \mathcal A[x(t)+q(t,a)]$) then the corresponding invariant is a conventional conserved quantity -- e.g. $x(t) \to x(t) + a$ ($y=\openone$, the ones-vector) generates the momentum, $x(t) \to x(t+a)$ ($y=\dot x$) generates the energy, etc.

  It is instructive to consider a case where the action is not invariant to the transformation.  For example, the form of the action is changed using the substitution for momentum in the simple harmonic oscillator ($L = m {\dot x}^2/2 - k x^2 / 2 \ne m {\dot x}^2/2 - k (x+a)^2 / 2$).  However, if we assume the existence of a generalized momentum, $\hat p$ that is nonetheless an invariant of some `complete' system, we can define (using \ref{eq:da})
\begin{equation*}
 \frac{d \hat p(t)}{dt} = \dot p - F = m\ddot x + kx
\end{equation*}
as a momentum {\em exchange}.  If the net momentum of the observed system changes by more than its internal force, it implies that an external system has lost exactly this amount of momentum.
In the presence of noise, this quantity may not be conserved ($d \hat p$ is stochastic) -- so that one effect of the noise is to re-distribute $\hat p$ over the system and the bath.

  In general, we can define a set of constraints on $\{\avg{dI_j(t)/dt}\}_{j=0}^m$ using an $N\times m+1$ matrix $Y$, whose columns correspond to the constraint directions $\eval{\pd{q_j(t,a)}{a}}{a=0}$ and whose leftmost column is reserved for the energy $Y_{*,0}=\dot x$.  In this case, the vector of exchanges for a given trajectory over a given time interval, $\epsilon$, is
\begin{equation}
\label{eq:dI}
dI(t) \equiv -Y(t)^T \dA{t} \epsilon
.
\end{equation}
In this equation, the presence of $\epsilon$ is used to implicitly denote the Stratonovich integral (see Appendix).

  These considerations have shown a simple method for including the influence of
an external system on the dynamics.  If we assume the existence of some
``total'' invariant between the system and the bath, then it makes sense to
enforce a stochastic constraint on the average change $\avg{dI(t)}$.  We should note our fixed sign convention, where $I$ is always taken to be a quantity belonging to the central system under consideration.
Before moving on to discuss the obvious connection of these changes to the
thermodynamic work, we shall first consider the significance of the Lagrange
multipliers in this formalism.

  Collecting constraints on $\dA{t}\dA{t'}\epsilon$, $dI$ and carrying out the maximum entropy
procedure specified in Eq.~\ref{eq:rSmax} for determining $\dA{t}$ for each time, $t$,
given a history $x(t')$ for $t'\le t-\epsilon, \epsilon\to0^+$,
\begin{align}
\begin{split}
 \Pr{\dA{t}\epsilon|x(t')_{t'=0}^{t-\epsilon}} &=
    \exp\left[-\epsilon\dA{t}^T G \dA{t} - \dA{t}^T \epsilon
      \int_0^{t-\epsilon} G_{t-\tau} \dA{\tau} d\tau + \epsilon\dA{t}^T Y(t) \beta/2
    \right]\\
     &\quad \times Z[\beta, x(t')_{t'=0}^{t-\epsilon}]^{-1}
\end{split} \notag \\
\begin{split}
\label{eq:trans}
  &= \exp\left[-(\dA{t}\epsilon-\mu(t))^T \frac{G}{\epsilon} (\dA{t}\epsilon-\mu(t))
         + \mu(t)^T \frac{G}{\epsilon} \mu(t)
    \right] \\
  &\quad \times Z[\beta,x(t')_{t'=0}^{t-\epsilon}]^{-1}
\end{split} \\
C &\equiv (2G)^{-1} \notag \\
\mu(t) &\equiv C \left( Y(t)\beta\epsilon/2
           - \epsilon \int_0^{t-\epsilon} G_{t-\tau} \dA{\tau} d\tau \right) \notag
.
\end{align}
In this equation, $Y\beta$ is a vector with the dimension of the system coordinates, since $\beta=[\beta_0,\ldots,\beta_m]^T$.  The action deviation, $\dA{t}\epsilon$, then follows a Normal distribution with mean $\mu(t)$ and single-time variance/covariance matrix $C\epsilon$.  The history integral in the above equation could alternatively have been written in terms of a time-dependent covariance function.  In this work, however, we will not be concerned with the calculation of history-dependent partition functionals for which this transformation becomes useful.

  Enforcing a constraint for the Hamiltonian energy change
$\avg{dH}=-\dA{t}^T \dot x\epsilon$ using the Lagrange multiplier $\beta_0/2$ (along with a possibly empty set of additional constraints in the form of Eq.~\ref{eq:dI}), leads directly to a Generalized Langevin equation
\begin{align}
\dA{t} \epsilon &= \mu(t) + \left(C \epsilon\right)^{1/2} z(t) \notag \\
\label{eq:langevin}
\dot p \epsilon &= F(x(t))\epsilon - C \left( \dot x \beta_0 \epsilon/2 + \tilde Y^T\beta\epsilon/2
           + \epsilon\int_0^{t-\epsilon} G_{t-\tau} (\dot p(\tau) - F(\tau)) d\tau \right)
	+ C^{1/2} \dW(t)
.
\end{align}
Here the centered (Stratonovich) Wiener process increment has been substituted for the standard normal random variate, $z(t)$, at time $t$ using $\dW(t)=\epsilon^{1/2}z(t)$.
It is well-known from the Fokker-Planck equation\cite{hrisk96}
\begin{equation*}
\pd{\rho(x,p)}{t} = -(M^{-1} p)^T \nabla_x \rho - \nabla_p^T \left[
  F\rho - C \dot x \beta_0 \rho/2 - C \nabla_p \rho/2 \right]
,
\end{equation*}
that the solution to this equation (in the memory-free case, but see also Ref.\citenum{tfran07}) is the canonical distribution with temperature $\beta^{-1}_0 = k_B T$ (where $k_B$ is the Boltzmann constant).
This solution is independent of $G$, suggesting a natural parametrization
for the Langevin equation is in terms of the temperature and $C^{1/2}$, related
to the thermal conductivity or rate of temperature equilibration (see Eq.~\ref{eq:temp}).
Generally, if an invariant, $I$, can be expressed as a function of $x,p$, then a result similar
to the above should hold for other common equilibrium thermodynamic ensembles as
well, such as the $N,P,T$ ensemble where $\avg{dV}$ constitutes an additional dynamic variable and
constraint.\cite{sfell95}
This establishes the physical interpretation of the Lagrange multipliers as the thermostatic variables of the bath that dictate the eventual equilibrium of the system.
Note that increasing $\beta$ tends to decrease $\avg{dI}$, for example increasing pressure will drive the volume downward.

  The present work shares some conceptual similarity to the second entropy of Attard~\cite{patta06,patta09}.  In this report, however, we have been able to derive our results in a mathematically rigorous way directly from two extremum principles, a maximum entropy expression for the transition probability (Eq.~\ref{eq:p}), and constraints derived from an action functional (Eq.~\ref{eq:dAdx}).  This allows for trivial generalizations to systems coupled with arbitrary reservoirs.  In addition, there is a clear physical motivation for the transition entropy and the full nonequilibrium entropy production which allows us to find the work done on the system by each constraint -- as will be shown in the next section.

  The above set of equations is also sufficient for defining nonequilibrium analogues of intensive thermodynamic 
variables such as the temperature.  This can be done by adding a hypothetical constraint, 
$\avg{dI_j}$, defined for some set of atoms or region of space in the system.  Analogous to the operation of a 
thermometer (zero energy exchange on imposing stochastic and damping terms) to define temperature,
we then require that no work is done on average, $\avg{dI_0}=0$.  Integrating using the Stratonovich rules developed in the appendix, we find
\begin{align}
\label{eq:temp}
 \avg{dI_0}/\epsilon &= \frac{1}{2}[\operatorname{Tr}(M^{-1} C) - \beta_0 \avg{\dot x_i^T C \dot x_i}] 
.
\end{align}
The result is the intuitive kinetic temperature, and is especially simple if we choose $C=2M\gamma/\beta_0$ as is common for the Langevin equation.  In that case, the ensemble average kinetic energy at each instant determines the temperature. For Boltzmann-distributed $\dot x$, regardless of the choice of $C$, the average heat flow is zero when $\beta_0$ determines the temperature.  If different types of particles can be coupled to separate thermostats, such as in plasmas, then it becomes physically meaningful to speak of separate ionic and electronic temperatures.


  To end this section, we show that it is possible to derive a Brownian limit from
our action functional approach using the alternative Lagrangian, Eqns.~\ref{eq:L1}-\ref{eq:dAdv}.
First, it can be shown that changes in the Hamiltonian are recovered by applying
Eq.~\ref{eq:dI}
\begin{align*}
-\dot q^T \tfrac{\delta\mathcal A}{\delta q(t)} - \dot v^T \tfrac{\delta\mathcal
A}{\delta v(t)}
 &= \dot q^T (\pd{U}{q} + M \dot v) - \dot v^T M (\dot q - v) \\
 &= \frac{d}{dt} [U(q) + v^T M v / 2]
\end{align*}
Next, applying Eq.~\ref{eq:trans} and assuming $\tfrac{\delta\mathcal A}{\delta q(t)}=0$ exactly,
we find a combined equation
\begin{align}
M \dot v \epsilon &= F(q(t))\epsilon \notag \\
\label{eq:brown}
\dot q \epsilon &= v\epsilon + C_q \left( \dot p \beta_0\epsilon/2
           - \int_0^{t-\epsilon} G_{q,t-\tau} (\dot q(\tau) - v(\tau)) d\tau
                      \right) + C_q^{1/2} \dW(t)
\end{align}
The second equation has the form of a Generalized Brownian motion equation, but includes
terms related to the process $v(t)$.  In particular, if the process $v(t)$ becomes unknown, then the
best guess form $M \dot v=F(x)$, and a streaming velocity $v = v_0$
generate an appealing equation for Brownian motion.  More rigorously, if the
process $v(t)$ is assumed to be unknown, updates $\dot q$ should be made based
on a stochastic realization of $v(t)$ whose average will generate the streaming
velocity $\avg{v(t)} = v_0$.

  To the best of the author's knowledge, this is a novel derivation of the Brownian limit that does not require an explicit limiting process of a infinitely massive particle, or infinite momentum jumps between position changes.  Instead, these two assumptions are implicitly present in assuming that $v,M\dot v$ are known during each position update.  The memory term derived here is similar to the form postulated in Ref.~\citenum{patta08}, which lead to a quantitative treatment of memory effects.  Here we can see it to be a natural consequence of placing constraints on the squared deviation of the action and the energy change at each time-step.

\section{ Irreversible Thermodynamics}
\label{sec:thermo}

  Having firmly established the connection to the equilibrium distribution above, we can construct a high-level view of any process employing a series of transition probabilities to effect a change in the state of the system.  This construction will lead naturally to a view of the process in terms of a thermodynamic path transforming one type of energy into another with a concomitant irreversible entropy production.

  To begin, we exactly define a system state, $A$, as any information which is known about a system that is sufficient to construct a probability distribution for its variables, $\Pr{x|A}$.  The machinery of statistical mechanics can then be used to propagate this information to system states at other time-points and under alternate possible processes.

  The work of Joule and Thomson showed that there exists a series of mechanical operations that can be performed to effect a transition between any two thermostatic states,\cite{callen} however this transition can only take place in the direction of increasing entropy.  The entropy increase comes about because of experimental inability to control the detailed motions of all particles, and is therefore zero in the case of completely controllable mechanical work.  Thus, it is important to define a mechanical, adiabatic process, in which all work is completely controlled by letting $C \to 0$ with constant external force experienced by the system $F^\text{sys}_{\text{ext},j} \equiv -\frac{C}{2} y_j \beta_j \to \pd{x(t)}{I_j(t)} F_{\text{ext},j}(t)$.
This is formally a zero-temperature, continuous-time limit, since at at a finite temperature there is some amount of uncertainty about the exact state of $I$ on short timescales, which leads to a discrepancy between the force exerted by the external system and its ``long-timescale'' counterpart experienced by the system.
In general, only a subset of work values can be controlled, and before proceeding it will be necessary to solidify the concept of controllable work.

  If this work is to be delivered by an external thermostatic system, for example an adiabatically coupled piston, then the first law of thermostatics gives $F_{\text{ext},j} = -dU_\text{ext}/dI_j = dU_\text{ext}/dI_{\text{res},j} = -\beta_j/\beta_0$ if the force can be assumed constant over a sufficiently short time-step.
Mechanically, this force corresponds to the force on a wall exerted by a spring placed externally to it.  The total force on the wall is, of course,
\begin{equation*}
 F_{\text{tot},j} = -\frac{dU_\text{int}+dU_\text{ext}}{dI_j} \equiv F_{\text{int},j} + F_{\text{ext},j}
.
\end{equation*}
Which implies that if a two spring system were disconnected after a change $d\tilde I \equiv [dI_1,\ldots,dI_m]^T$, their internal potential energies would have changed by an amount
\begin{align}
 dU_\text{int} &= -d\tilde I^T F_\text{int} \notag \\
\label{eq:dUe}
 dU_\text{ext} &= -d\tilde I^T F_\text{ext} = d\tilde I^T \tilde\beta/\beta_0 \\
\tilde\beta &\equiv [\beta_1,\ldots,\beta_m]^T \notag
.
\end{align}
The sum of these two energy changes is not necessarily zero due to the possibility of momentum change.  Using the known energy change of the system, it should then be possible to solve for the change in kinetic energy of the constraint.  For the system, the {\em total} energy change is given by
\begin{align}
 dE &= -\tfrac{\delta \mathcal A}{\delta x}^T \dot x \epsilon = dI_0 \notag \\
    &= dW + dQ \notag \\
 dW_j &\equiv -\tfrac{\delta \mathcal A}{\delta x}^T y_j \dot a_j \epsilon \notag \\
\label{eq:dW}
 dW &= d\tilde I^T \dot a \\
\label{eq:dQ}
 dQ &= -\left(\dot x - \tilde Y \dot a \right)^T \tfrac{\delta \mathcal A}{\delta x} \epsilon
,
\end{align}
where we have used Eq.~\ref{eq:da} for $\pd{\mathcal A}{a}$, multiplied by $da/dt$, to define work values and a corresponding ``non-mechanical'' energy transfer, $dQ$.  In the adiabatic limit, all energy transferred to the system by external forces should be reflected by known mechanical changes (related to $\{I\}$) -- which is precisely what $\dot a$ allows us to do.  Note also that unless specifically denoted `ext' all quantities refer to the central system, so that $dW$ means the work done on the system.

  In the mechanical limit, $-\dA{t} = \dot p - F_\text{int} = F^\text{sys}_\text{ext}$, the external force experienced by the system in the absence of the thermostatting random noise and $Y_{*,0}=\dot x$.  Substituting this quantity from the Langevin equation (\ref{eq:langevin}),
an adiabatic, mechanical system must satisfy
\begin{align}
 dQ &= -(\dot x - \tilde Y \dot a)^T \tfrac{\delta \mathcal A}{\delta x} \epsilon \notag \\
\label{eq:dQmech}
    &= -\frac{1}{2} (\dot x - \tilde Y \dot a)^T C \tilde Y \tilde \beta \epsilon = 0
.
\end{align}
The last section gave some physical insight into the quantities $\dot a$.  A mathematical consideration of the previous equation shows that $\tilde Y \dot a \equiv \sum_{j=1}^m y_j \dot a_j $ ($\tilde Y$ being identical to $Y$ with the first column removed) can be understood as a projection, removing components of $\dot x$ parallel to $\delta\mathcal A/\delta x$.  If we therefore define (writing the Moore-Penrose pseudoinverse of $A$ as $A^+$)
\begin{equation}
\label{eq:dota}
\dot a \equiv (C^{-1/2} \tilde Y)^+ C^{-1/2} \dot x
,
\end{equation}
then $dQ=0$ identically.  For an example application, $y=\openone$ (ones-vector) is associated with the system momentum, and Eq.~\ref{eq:dota} generates the average velocity $\dot a = \openone^T \dot x/N$ when $C=cI$.  Eq.~\ref{eq:dota} is invariant to multiplication of $C$ by a constant, and so persists in the deterministic limit.  The work done on the system (Eq.~\ref{eq:dW}) is
\begin{equation*}
 dW = -\frac{\delta\mathcal A}{\delta x}^T \tilde Y (C^{-1/2} \tilde Y)^+ C^{-1/2} \dot x \epsilon
.
\end{equation*}

  Note the information-theoretic quality of the work defined by the above equation.  If separate reservoirs existed that were able to independently influence the motion of each particle in the system, then $\tilde Y$ would become an identity matrix, and $d\mathcal W$ would equal $dE$ identically.  In the presence of noise, the work done does not necessarily equal the energy change.  In the $y=\openone$ example above, the work is
\begin{equation}
\label{eq:pdiss}
\tfrac{1}{N} (\sum_a \dot p_a-F_\text{int,a})(\sum_a \dot x_a) \epsilon
.
\end{equation}
Here ``$\sum_a$'' denotes a sum over $N$ (one-dimensional) atoms, with obvious extension to multiple dimensions. Because this interaction controls only the total system momentum, the work is computed using the average velocity change.  As shown in the appendix, the expectation of this stochastic integral is $\avg{dW_p}/\epsilon = \beta_0/2 \dot x_a^T C (\dot x_a+\lambda/\beta_0)$.  Similarly, an electric field can couple only to the net dipole moment of a system.  This implies the transformation of applied energy to heat if it cannot be manipulated to perfectly match fluctuations in the driven variable.

  The kinetic energy change of $I$, ascribed to the reservoir, can be determined in the mechanical limit from
\begin{align}
  0 &= dU_\text{ext} + dK - dW \notag \\
  \Rightarrow
\label{eq:dK}
  dK &= -dI^T (\tilde\beta/\beta_0-\dot a)
,
\end{align}
using Eqns.~\ref{eq:dUe} and~\ref{eq:dW}\footnote{Note the similarity of this form to $\int \dot p - F \; dt$.}.  The above equations thus completely describe any exchange of mechanical energy between deterministic systems exerting known forces.  If the two-spring system considered above were disconnected at time $S$, an outside observer absorbing the kinetic energy of the wall (Eq.~\ref{eq:dK}), the energy change of the reservoir would reduce to the usual thermostatic potential change $\int_0^S dU_\text{ext}(t)$.

  A well known consequence of Liouville's theorem is that the entropy change is zero in a completely deterministic process.\cite{brobe71,brobe93}  Using adiabatic processes, then, it is possible to propagate a starting state, $A$, to any state with constant entropy.
  If, however, phase space volume were not preserved, it would amount to a discarding of information on the state of the system at a given time (for example, by integrating the probability over short time intervals).  Then the amount of work that can be recovered from the system will become less than the amount input.  In an extreme case, all information about the system may have been lost, flushing the corresponding information content to zero.  Starting from this unknown state, $p^0$, the probability of a given frequency distribution, $p$ is approached by $P(p) \propto e^{\H[p]}$, with $\H[p]$ the familiar information entropy functional
\begin{equation}
 \label{eq:S}
\H[p] = -\int p_i \ln p_i/p^0_i
.
\end{equation}
Further, if exchanges of conserved quantities during some process $A\to B$ are known, then the set $I(B)$ are also known from $A$, and this information can be usefully employed to increase the amount of work that can be recovered -- showing the entropy as a measure of ``lost work''.  It therefore stands to reason that any adiabatic process should be described by not only the above mechanical work values, but also the change in information entropy due to information loss.

  Next, consider allowing heat exchange in addition to controllable work.  Assuming an infinitely large reservoir (or a short enough time-step), added heat will cause a negligible change in reservoir temperature.  Because we have assumed the work done on each reservoir (\ref{eq:dUe}) can be reversibly stored, these are not associated with an entropy change.  We therefore introduce the physical entropy change in the reservoir as due only to exchange of heat, or ``non-work'' energy, $dS_\text{ext} = \beta_0 dQ_\text{ext}$, with $\beta_0^{-1}=k_B T_\text{ext}$, and $dQ_\text{ext}$ originating from heat removed from the system plus a kinetic energy, $dK$, assumed to be recoverable only as heat.
In this work, different notations are used for the information entropy, $\H$, the physical entropy, $S$, the caliber, $\sigma$, and the caliber-like functional, $\sigma^*$.  All of these quantities are defined to be unitless and have some relation to the information entropy of Eq.~\ref{eq:S}.
\begin{align}
dQ_\text{ext} &= -dQ + dK = -dE - dU_\text{ext} \notag \\
\label{eq:dQr}
  &= \dA{t}^T {\dot x}\epsilon - {d\tilde I}^T \tilde\beta/\beta_0 = -dI^T\beta/\beta_0
.
\end{align}
The energy rejected to the reservoir as heat can alternately be understood as the total energy dumped to the environment minus the energy removed ``reversibly'' ($dU_\text{ext}$).  This interpretation shows that if some of the changes in the environment were re-classified as irreversibly stored, so that the information $\beta_j dI_j$ becomes lost, then this is equivalent to adding that energy to the total $dQ_\text{ext}$.
The total energy rejected to the environment is recoverable (the mechanical limit considered above) if and only if $dQ_\text{ext}=0$, implying $dE = dI_0 = -d\tilde I^T\beta/\beta_0$.  These considerations again highlight the subjective nature inherent in the definition of irreversibility.

  Connecting back to the usual thermostatics, $U_\text{ext}$ plays the role of an energy for the reservoir.
\begin{align*}
 dU_\text{ext} &= dE_\text{ext}-\beta_0^{-1} d S_\text{ext} \\
 &= d\tilde I^T\tilde\beta/\beta_0
\end{align*}
Where we must provide an experimental justification for the ability to use or store the energy terms appearing in the above sum.

  For any transformation, $A\to B$, the total entropy change deriving from information loss and interactions with the environment is given by the change in information entropy plus the heat exchange term
\begin{equation}
\label{eq:ent1}
\Delta S_\text{inf,tot} = \H[P(x|B)]-\H[P(x|A)] + \beta_0 Q_\text{ext}
.
\end{equation}
It should be noted that this formula is still not complete if there is a change in the phase space between $A$ and $B$, for example if particles are added/removed, or if the state space is uniformly dilated.  In this case, we have extra information on the region of phase space occupied after a transition.  In general, if the state at time $i$ is known to be $x_i$, then the size of the region of configuration space accessible at time $i+1$ is $Z_i(\beta,x_i)$ (Eq.~\ref{eq:Z}).  This reduces the entropy at $i+1$ to $S_{i+1} - \ln Z_i[\beta,x_i]$.  Writing this down for each transition,
\begin{align}
\Delta S_\text{tot} &= \avg{\sum_{i=0}^{S-1} dS_i} \notag \\
\label{eq:ent}
dS_i &\equiv -\ln \frac{\Pri{i+1}{x_{i+1}}Z_i[\beta,x_i]}{\Pri{i}{x_i}Z_i[-\beta,x_{i+1}]} - dI^T \beta
.
\end{align}
The above equation has been symmetrized by including the corresponding entropy decrease for $i$ given that $x_i$ was inferred from $x_{i+1}$, using the forward step probability but with reversed forces, $-\beta$.  Support for this form is given in the next section, where it is also shown that $\Delta S_\text{tot} \ge 0$ using the Gibbs inequality.

  Nevertheless, Eq.~\ref{eq:ent1} can already be applied to Langevin and Brownian motion with uniform diffusion constants.  As another example, applying Eq.~\ref{eq:ent1} to a process taking any point to the equilibrium distribution shows that the expected entropy change for this process is the usual system entropy difference plus $\beta_0 Q_\text{ext} = -\beta_0 (\avg{E|B}-\avg{E|A})-\Delta \tilde I^T\tilde\beta = (I(A)-I(B))^T \beta$.  Because the end-point entropies are fixed, alternate processes for transforming $A\to B$ are restricted to varying $\Delta S_\text{ext}=\int \beta_0 dQ_\text{ext}$.
For such processes, we may write a strong form for the Clausius form of the second law
\begin{equation}
\Delta S_\text{tot} = \Delta\H + \int \beta_0 dQ_\text{ext} \ge 0
.
\end{equation}
Here, $\Delta \H$ is a function of the end-points, and $\beta_0$ and $dQ_\text{ext}$ are fully variable along the path.
Choosing a path from a fully specified distribution, $A$, to a maximum entropy distribution, $I(A)$, and then to the ending maximum entropy distribution $I(B)$, we may employ a quasistatic, ``reversible,'' path between the two maximum entropy distributions, so that $\min \int \beta_0 dQ_\text{ext}(A\to I(B)) = \min \int \beta_0 dQ_\text{ext}(A\to I(A))$.  The heat evolved in this best-case process has its origin in the re-classification of information that occurs during the coarsening of $A$, in accordance with conclusions on Maxwell's demon.\cite{hleff03}

  The equilibrium theory is therefore contained in the present development in the form of slow, quickly relaxing processes.  This perspective shows the intimate connection between coarse-graining that assumes infinitely fast relaxation of the reservoir and the traditional theory of quasi-static processes.  However, the complete theory also permits an investigation of both relaxation processes and entropy production in time and history-dependent processes moved arbitrarily far from equilibrium by coupling to simple thermostatic reservoir systems.

\section{ Predictive Statistical Thermodynamics}
\label{sec:pred}

  As noted in the introduction, there is a fundamental difference between applying maximum entropy to a complete trajectory and to each transition probability distribution individually.  Referring to Eqns.~\ref{eq:p} and~\ref{eq:Z}, the probability for a path $\Gamma\equiv\{x_i\}_0^S$ (using a starting distribution $\Pr{x_0}$ on $x_0$) is
\begin{align*}
 \Pr{\Gamma} &= \prod_{i=0}^{S-1} \Pr{x_{i+1}|\{x\}_0^i} \Pr{x_0} \\
   &= \Pri{0}{\Gamma|x_0} \frac{e^{-\eta[\Gamma;\{\lambda\}]}}{\mathcal Z[\{\lambda\},\{x\}_0^{S-1}]}  \Pr{x_0} \\
 \eta[\Gamma;\{\lambda\}] &\equiv \sum_{i=0}^{S-1} \sum_k \lambda_{k,i} f_k(x_{i+1},\{x\}_0^i) \\
 \mathcal Z[\{\lambda\},\{x\}_0^{S-1}] &\equiv \prod_{i=0}^{S-1} Z[\lambda_i,\{x\}_0^i] \\
 \Pri{0}{\Gamma|x_0} &\equiv \prod_{i=0}^{S-1} \Pri{0}{x_{i+1}|\{x\}_0^i}
.
\end{align*}
Defining a path information entropy, or caliber, associated with the number of possible ways that a given $\Pr{\Gamma}$ could be observed, we have
\begin{align}
\sigma_\Gamma &\equiv -\int \Pr{\Gamma} \ln \frac{\Pr{\Gamma}}{\Pri{0}{\Gamma}} \; d\Gamma \notag \\
  &= \int \Pr{\Gamma} (\eta + \ln \mathcal Z - \ln \frac{\Pr{x_0}}{\Pri{0}{x_0}}) \; d\Gamma \notag \\
\label{eq:caliber}
  &= \avg{\eta} + \avg{\ln \mathcal Z} + \H_0
.
\end{align}

  This equation shows that the path entropy is an average of maximum entropy increments given by Eq.~\ref{eq:rS}.  This form telescopes in the following way.  Suppose only the process from time $0$ to $j \le S$ is of interest.  In this case, the above entropy expression can be separated from
\begin{align*}
\sigma_\Gamma &= \sum_{i=j}^{S-1} \avg{\sum_k \lambda_{k,i} f_k(x_{i+1},\{x\}_0^i)}
                         + \avg{\ln Z[\lambda_i,\{x\}_0^i]} + \sigma_{\Gamma'} \\
\intertext{with the non-anticipating partial sum}
\sigma_{\Gamma'} &= \sum_{i=0}^{j-1} \avg{\sum_k \lambda_{k,i} f_k(x_{i+1},\{x\}_0^i)}
                         + \avg{\ln Z[\lambda_i,\{x\}_0^i]} + S_0
.
\end{align*}
For times between $0\le j\le S$, the first line above has exactly the same form as Eq.~\ref{eq:caliber}, with $\H_0$ replaced by $\sigma_{\Gamma'}$.  That is, the starting distribution has become a multi-step probability distribution with no change in the total caliber.

\subsection{ Path Averages and Fluctuations}
  Analogous to the equilibrium theory, we should expect that a cumulant expansion of an appropriate path free energy will yield path averages, fluctuations, etc.  From Eq.~\ref{eq:caliber}, a path free energy functional can be defined as
\begin{equation}
\label{eq:pathF}
\mathcal F[\lambda] \equiv -\avg{\ln\mathcal Z} = \avg{\eta} - \sigma_\Gamma + \H_0
.
\end{equation}
  Expanding
\begin{align*}
\pd{\mathcal F}{\lambda_{k,i}} &= \pd{}{\lambda_{k,i}} \int \left(-\sum_{j=0}^{S-1}
     \ln Z[\lambda_j;\Gamma]\right) \Pr{\Gamma} \; d\Gamma \\
  &= \avg{f_k(x_{i+1},\{x\}_0^i)}
  - \int \ln \mathcal Z \pd{}{\lambda_{i,k}} \Pr{\Gamma} \; d\Gamma \\
  &= \avg{f_k(x_{i+1},\{x\}_0^i)} + \avg{f_k(x_{i+1},\{x\}_0^i) - \avg{f_k(x_{i+1},\{x\}_0^i)}} \\
  &= \avg{f_k(x_{i+1},\{x\}_0^i)}
.
\end{align*}
we see that the derivatives of $\mathcal F$ indeed give the path averages, $\avg{f}$.

  This result is valid for any chosen set of $\lambda$ and corresponding functions, $f_k$.  It is therefore possible to formally use the above to compute arbitrary path expectations, even if they do not affect the dynamics ($\lambda=0$).  We can also use the above equation to formulate a first law of time-dependent, nonequilibrium thermodynamics
\begin{equation}
\label{eq:one}
d\mathcal F[\{\lambda\}_0^{S-1}] = \sum_{i=0}^{S-1} \sum_k \avg{f_k(x_{i+1},\{x\}_0^i)|\lambda} d\lambda_{k,i}
.
\end{equation}

  We can similarly compute second derivatives to give
\begin{align}
 \frac{\partial^2 \mathcal F}{\partial \lambda_{k,i} \partial \lambda_{l,j}} &= 
    \avg{f_{k,i}}\avg{f_{l,j}} - \avg{f_{k,i} f_{l,j}} \notag \\
\label{eq:gibbs}
  &\equiv -\operatorname{Cov}\left[ f_k(x_{i+1},\{x\}_0^i), f_l(x_{j+1},\{x\}_0^j) \right]
.
\end{align}
Using the Legendre transformations ($\mathcal F - \lambda_h\pd{\mathcal F}{\lambda_h}$) given in Eq.ns~\ref{eq:F}-\ref{eq:dF}, these fluctuations can be transformed to ensembles with constrained averages (acceleration, particle flux, etc.), $\avg{f}$, rather than thermodynamic forces, $\lambda$.

  A simple derivation for such equations can be given following Ref.~\citenum{callen}.  Performing a second-order expansion using Eq.~\ref{eq:gibbs}, and writing the result in matrix form,
\begin{align*}
 \begin{bmatrix}d\avg{f}_1 \\
  d\avg{f}_2 \end{bmatrix}
   &= \begin{bmatrix}
\frac{\partial^2 \mathcal F}{\partial \lambda_1 \partial \lambda_1}
    & \frac{\partial^2 \mathcal F}{\partial \lambda_1 \partial \lambda_2} \\
\frac{\partial^2 \mathcal F}{\partial \lambda_2 \partial \lambda_1}
    & \frac{\partial^2 \mathcal F}{\partial \lambda_2 \partial \lambda_2}
\end{bmatrix} \begin{bmatrix}d\lambda_1 \\
d\lambda_2 \end{bmatrix} \\
 &\equiv \begin{bmatrix}A & B \\ C & D\end{bmatrix} \begin{bmatrix}d\lambda_1 \\
d\lambda_2 \end{bmatrix} 
,
\end{align*}
we swap the sides of a set the averages $d\avg{f}_2$ and their corresponding forces $d\lambda_2$.
\begin{equation*}
  \begin{bmatrix}I \\ 0\end{bmatrix} d\avg{f}_1
- \begin{bmatrix}B \\ D\end{bmatrix} d\lambda_2
= \begin{bmatrix}A \\ C\end{bmatrix} d\lambda_1
- \begin{bmatrix}0 \\ I\end{bmatrix} d\avg{f}_2
\end{equation*}
Re-assembling the matrices on each side, and inverting $\begin{bmatrix}I & -B \\ 0 & -D\end{bmatrix}$,
\begin{align}
 \begin{bmatrix}d\avg{f}_1 \\
  d\lambda_2 \end{bmatrix}
   &= \begin{bmatrix}
A - B D^{-1} C & - B D^{-1} \\
- D^{-1} C & D^{-1}
\end{bmatrix} \begin{bmatrix}d\lambda_1 \\
d\avg{f}_2 \end{bmatrix} \notag \\
\label{eq:Gibbs}
   &= \begin{bmatrix}
\frac{\partial^2 \Psi}{\partial \lambda_1^2}
    & \frac{\partial^2 \Psi}{\partial \lambda_1 \partial \avg{f}_2} \\
-\frac{\partial^2 \Psi}{\partial \avg{f}_2 \partial \lambda_1}
    & -\frac{\partial^2 \Psi}{\partial \avg{f}_2^2}
\end{bmatrix} \begin{bmatrix}d\lambda_1 \\
d\avg{f}_2 \end{bmatrix} \\
\Psi &\equiv \mathcal F - \lambda_2^T\avg{f}_2,\quad d\Psi = \avg{f}_1^T d\lambda_1 - \lambda_2^Td\avg{f}_2
.
\end{align}

  The above manipulation, well-known in the theory of linear, local equilibrium processes,\cite{skjel08} are seen as Gibbs relations in the general nonequilibrium theory developed here.  These equations describe a change of ensemble in the equilibrium theory~\citenum{jlebo67}.  In this respect, they form a basis for connecting stochastic, Langevin dynamics simulations (e.g. Eq.~\ref{eq:langevin}) to constant kinetic energy, solute flux, etc. ensembles studied extensively in nonequilibrium molecular dynamics simulations\cite{devan08}.
In addition, the upper-right matrix element, $d\avg{f}_1 = - \operatorname{Cov}[f_1,f_2] {\operatorname{Cov}[f_2,f_2]}^{-1} d\avg{f}_2$ appears as the starting point for the Mori projector-operator method\cite{hmori65} as has been discussed by Jaynes.\cite{ejayn79,ejayn80}

  Using the action-type conserved quantity constraints (Eq.~\ref{eq:dI}) in the free energy functional, its derivatives generate the fluxes $\frac{\delta \mathcal F}{\delta (\beta_j(t)/2)} = \avg{dI_j(t)}$.  At equilibrium, these fluxes are zero.  Given information about the previous history of the system, $\avg{dI(t-\tau)}$, we can use Eq.~\ref{eq:Gibbs} to find the linear change
\begin{align}
\pd{\Psi}{(\beta(t)/2)} &\approx \sum_{t'<t} \frac{\partial^2 \Psi}{\partial(\beta(t)/2)\partial(\beta(t')/2)} \avg{dI(t')} \\
 &= -\avg{dI(t+1)dI(t)^T}\avg{dI(t)dI(t)^T}^{-1} \avg{dI(t-1)}
\equiv -\frac{\sigma_1}{\sigma_0} \avg{dI(t-1)}, \quad \text{one step} \\
 &= -\sigma_1 {\begin{bmatrix}\sigma_0 & \sigma_1 \\ \sigma_1 & \sigma_0 \end{bmatrix}}^{-1}
     \begin{bmatrix}\avg{dI(t-1)}\\ \avg{dI(t-2)}\end{bmatrix}, \quad \text{two steps}
\end{align}
This is the linear equation of motion for a near-equilibrium system under a mechanical driving force, and explains the ubiquitous use of Fourier transforms in solving these equations.
It could be expanded to arbitrary order using a Taylor series in $\pd{\Psi}{(\beta_j(t)/2)}$.

  The case of driving by thermal, or constant external force, conditions is much simpler, and we may accordingly treat the more complicated case of a transient process.  According to Eqns.~\ref{eq:one} and~\ref{eq:gibbs}, we may expand about an arbitrary initial distribution plus some reference program, $\{\beta\}_0^{S-1}$ (denoted by the zero subscript), to find the equation of motion for thermal driving.
\begin{align*}
\avg{dI(t)} &\approx \avg{dI(t)}_0 + \sum_{t'<t} \frac{\partial^2 \mathcal F}{\partial (\beta(t)/2)\partial(\beta(t')/2)} \delta(\beta(t')/2) \\
 &= \avg{dI(t)}_0 - \tfrac{1}{2} \sum_{t'<t} \avg{dI(t)dI(t')^T}_0 \delta \beta(t')
\end{align*}
This relation contains a factor of $1/2$, as in a derivation by Searles and Evans~\cite{dsear00} where it was shown to reduce to the Green-Kubo expression in the zero-field limit.  As noted in that work, the last term on its own is incorrect when a driving force is present.

  The leading term in the above is the flux in the reference process.  Choosing this reference process as a conducting steady-state shows one mechanism for the failure of the Onsager reciprocity relations.  The relations $B^T=C$ should hold in any case, but are only derivatives of the flux/force curve.  These derivatives are analogous to the fluctuation moments of the canonical ensemble, which can approximate the energy at a slightly altered temperature.  They are expansions about a fully nonlinear free energy functional, $\mathcal F$.

\subsection{ Path Perturbation and Connection to Entropy Production Theorems}

  The linear relations derived in the last section should not be expected to hold for large deviations in the nonequilibrium forces.  We can progress beyond this limitation by analogy to the transition from thermodynamic integration to free energy perturbation in equilibrium free energy calculations.
Any two processes on the same path space, $\{\Gamma\}$, can be compared using a likelihood,
\begin{equation}
\label{eq:l}
e^{l_{A\to B}[\Gamma]} \equiv \frac{\Pri{B}{\Gamma}}{\Pri{A}{\Gamma}}
 = e^{-(\eta_B-\eta_A)} \frac{\mathcal Z_A[\Gamma]}{\mathcal Z_B[\Gamma]}
.
\end{equation}
The likelihood obeys $\avg{e^{l_{A\to B}[\Gamma]}}_A = 1$, and (by the Gibbs inequality) $\avg{l_{A\to B}[\Gamma]}_A \le 0 \le \avg{l_{A\to B}}_B$.  The distribution of the weight satisfies the perturbation formula\cite{cbenn76}
\begin{align*}
 e^{L} \Pri{A}{L=l_{A\to B}[\Gamma]}
 &= \int e^{l_{A\to B}[\Gamma]} \delta(L-l[\Gamma]) \Pri{A}{\Gamma} \; d\Gamma \\
 &= \int \delta(L-l[\Gamma]) \Pri{B}{\Gamma} \; d\Gamma \\
 &= \Pri{B}{L=l_{A\to B}[\Gamma]}
.
\end{align*}

  It is possible to express path averages using the above quantity as
\begin{equation}
\label{eq:rewt}
\avg{a[\Gamma]}_B = \avg{a[\Gamma]e^{l_{A\to B}[\Gamma]}}_A
.
\end{equation}
However, in the equilibrium theory, a constant related to the free energy difference is usually cancelled on the right-hand side of this expression.   There is no such constant in the above equation because we have not identified an appropriate extensive variable.  In the present case, we can use the free energy (Eq.~\ref{eq:pathF}) to define
\begin{align*}
e^{l'_{A\to B}[\Gamma]} &\equiv \frac{\Pri{B}{\Gamma} e^{-\mathcal F_B[\lambda]}}
  {\Pri{A}{\Gamma} e^{-\mathcal F_A[\lambda]}} \\
\intertext{in terms of which}
\avg{e^{l'_{A\to B}[\Gamma]}}_A &= e^{-(\mathcal F_B[\lambda]-\mathcal F_A[\lambda])}\\
\intertext{and}
\avg{x[\Gamma]}_B &= \frac{\avg{x[\Gamma]e^{l'_{A\to B}[\Gamma]}}_A}{\avg{e^{l'_{A\to B}[\Gamma]}}_A}
.
\end{align*}
Or course, the above expressions may be of little analytical use because $\avg{\ln Z[\lambda_i,\{x\}_0^i]}$ has simply been subtracted from $\eta[\Gamma]$ (in the exponent of $e^{\ln Z[\lambda_i,\{x\}_0^i]}$.  However, the transition probabilities usually adopt particularly simple forms, e.g. normal distributions.  If, for example, the transition probability takes every point to an equilibrium distribution, then the above expansion is particularly useful.

  It has been claimed that entropy production can be gauged by the ratio of forward to reverse path probabilities.\cite{gcroo99,cmaes99}  For the process derived in Sec.~\ref{sec:dyn}, we can define a reverse process by inverting the sign of the generalized forces, $\beta_i$, and normalizing the distribution separately for each $x_{i+1}$.  Such a reversal corresponds to an attempt at guessing whether energy has been added or subtracted during a step $i\to i+1$, with the action deviation constraint, $G$, unchanged.
Thus, we can define a ratio
\begin{align}
\label{eq:FT}
e^{dS_i} &\equiv \frac{\Pr{x_{i+1}|x_i    , \beta} \Pr{x_i}}
                      {\Pr{x_i    |x_{i+1},-\beta} \Pr{x_{i+1}}} \\
 &= \frac{\Pr{x_i}     p_0(x_{i+1}|x_i) Z[-\beta,x_{i+1}]}
         {\Pr{x_{i+1}} p_0(x_i|x_{i+1}) Z[ \beta,x_i]} e^{-dI^T \beta} \notag
.
\end{align}
Since $\frac{p_0(x_{i+1}|x_i)}{p_0(x_i|x_{i+1})} = \frac{p_0(x_{i+1})}{p_0(x_i)}$ by Bayes' theorem, this result exactly matches Eq.~\ref{eq:ent} arrived at through thermodynamic reasoning.

  Moreover, several steps can be concatenated to give
\begin{equation}
\label{eq:two}
 e^{\sum_{i=0}^{S-1} dS_i} = \frac{\Pr{\Gamma|x_0, \beta} \Pr{x_0}/p_0(x_0)}
                                  {\Pr{\Gamma|x_S,-\beta} \Pr{x_S}/p_0(x_S)}
.
\end{equation}
The above equation can therefore be viewed as a statistical basis for the second law of thermodynamics.  It is physically motivated by observing that the entropy increase is attributed to a combination of environmental entropy changes, $dQ_\text{ext}$, and information-like entropy changes,
\begin{equation}
-\ln \frac{\Pr{x_{i+1}} Z[ \beta,  x_i   ]/p_0(x_{i+1})}
              {\Pr{x_i} Z[-\beta, x_{i+1}]/p_0(x_{i})} \\
\end{equation}

  The above term implies an extra contribution to the total entropy change beyond Eq.~\ref{eq:ent1}.  To understand this contribution, consider a simple one-state system, to be transformed into a two-state system through the (unbiased) transition probability $\Pr{x_{i+1}|x_i} = p_0(x_{i+1}) = 1/2$.  The presence of the normalization in Eq.~\ref{eq:ent} leads to an additional factor of $-\ln Z[x_i]/p_0(x_{i+1}) = -\ln 2$ in the entropy.  However, this term is canceled by the entropy of the resulting state, $\Pr{x_{i+1}}=1/2$, so that the total entropy change for this process is zero.  If, instead, we perform a bit-set operation by going in the opposite direction, the entropy will increase if $\Pr{x_{i+1}}$ is any non-uniform state -- corresponding to our loss of information on discarding the bit.  A term of this form is exactly what we should have expected when writing down Eq.~\ref{eq:ent1}.  Applying this equation to a situation where a particle is added to the system, we find that the re-normalization will physically compensate for the expansion of phase space which leads to a difference in entropy measures.  Mathematically, this implies that the ``default'' measure $p_0(x_{i+1})$ is replaced by the re-normalized measure $p_0(x_{i+1}) Z^*_\text{tr}(x_{i+1})/Z_\text{tr}(x_{i})$ if we have information about the previous
state when calculating the entropy at state $i+1$.

  Equation~\ref{eq:two} is connected to the likelihood definition (\ref{eq:l}).  Defining $B$ as a ``forward'' process, starting from $\Pr{x_0}$ and employing $\lambda_B = \{G,\beta_{i,k}/2\}$, and $A$ as a ``reverse'' process, employing $\lambda_A = \{G,-\beta_{i,k}/2\}$.  Because
\begin{equation*}
e^{l_{A\to B}} = \frac{\Pr{\Gamma|B}}{\Pr{\Gamma|A}}
               = \frac{\Pr{\Gamma|x_0,B}\Pr{x_0|B}}{\Pr{\Gamma|x_S,A}\Pr{x_S|A}}
\end{equation*}
and the end-point distributions $\Pr{x_0},\Pr{x_S}$ have been pre-determined, we can write Eq.~\ref{eq:ent} as
\begin{align}
\label{eq:divS}
 \Delta S_\text{tot} &= \avg{l_{A\to B}}_B \\
\intertext{which implies}
\Delta S_\text{tot} \ge 0 \notag
.
\end{align}
This completes the connection between the physical entropy production defined in Sec.~\ref{sec:thermo} and fluctuation theorems of the form (\ref{eq:l}).

  However, the entropy production has not yet been connected to the caliber functional (Eq.~\ref{eq:caliber}).  To do this, we decompose Eq.~\ref{eq:FT} into the form of Eq.~\ref{eq:caliber} as
\begin{align}
\label{eq:Sconn}
\Delta S_\text{tot} &= \sigma^*_\Gamma - \sigma_\Gamma \\
\intertext{by defining}
\label{eq:rcaliber}
 \sigma^*_\Gamma &\equiv -\avg{\sum dI^T \beta/2}_B + \avg{\ln \mathcal Z[-\beta]}_B + \H_S
.
\end{align}
Although this definition is similar to the caliber, it is not an information entropy, since the averages are taken with respect to the forward probability distribution, $\Pr{\Gamma|x_0,\beta}\Pr{x_0}$.  The choice of the forward direction corresponds to the direction in which information propagates~\cite{ejayn57a}.  It determines the target distribution for taking the divergence (Eq.~\ref{eq:divS}).

\subsection{ Jarzynski's Equality}

  In the special case where a stationary distribution is known during each time-propagation step, a set of useful equalities can be derived simply from the re-weighting equation
\begin{equation}
 \avg{a(x(T))|\text{ss},T} = \avg{a(x(T))
  \prod_{t=1}^{T} \frac{\Pri{\text{ss}}{x(t),t}}
                       {\Pri{\text{ss}}{x(t),t-1}} | x_0\sim \text{ss},0}
.
\end{equation}
The first average represents a single-time average over the steady state distribution used to propagate the system during the transition $T\to T+1$.  The second average is a path average over stochastic trajectories beginning in the steady-state at time zero.

  The proof\cite{cjarz97} is by recursion from the property of the stationary distribution under time-propagation,
\begin{equation}
 \Pri{\text{ss}}{x(t+1),t} = \int  \Pri{\text{tr}}{x(t+1)|x(t),t} \Pri{\text{ss}}{x(t),t} \;dx(t)
.
\end{equation}
These relations will work for any stationary distribution of the transition probability.  Equality of the starting and ending temperatures is not required.

\section{ Connecting the First and Second Laws}

  In order to give concrete examples of Eqns.~\ref{eq:gibbs} and Eq.~\ref{eq:ent}, we must choose a set of coarse variables of interest ($\avg{f}$) and follow their time-evolution, $\avg{df(t)}$.  Most mesoscopic models contain hydrodynamic equations of motion for the solution density.  A rigorous route\cite{rleho10} to their derivation is by forming suitable integrals of the probability distribution function appearing inside the Liouville equation, and much of the early literature on nonequilibrium problems is focused on this derivation.  In the maximum transition entropy context, the resulting equations describe the propagation of a state of knowledge forward in time using an exact equation of motion.  The exact internal and external forces on the system are required using this route, and the projector-operator formalism is used to add the uncertainty introduced by mixing processes occurring below the size and time-scale of the density function.

  In the stochastic formalism developed here, the Liouville equation has been replaced with the Klein-Kramers and Smoluchowski equations.  In addition to convection, these equations also describe diffusion of probability that occurs because of loss of information during mixing processes.  Summing the Fokker-Planck equation for Eq.~\ref{eq:brown} over particles of each species type, $\alpha$, and integrating over coordinates other than that of a single, distinguished molecule (using the relations from Ref.~\citenum{rleho10}) gives the one-particle evolution equation
\begin{equation}
\label{eq:diffn}
 \pd{\rho_\alpha(r,t)}{t} = -\nabla_r \left(D(\beta F\rho(r,t)-\nabla_r\rho(r,t))\right)
.
\end{equation}
In this process, the force and diffusion coefficients have become averages over the probability distributions of the molecules of each type -- i.e. a potential of mean force.
Here, $D$ has been substituted for $C_x/2$.

  The form of equation~\ref{eq:diffn} corresponds exactly to the usual diffusion equation used for the continuum formulation of electrodiffusion equations~\cite{gkarn05}, without the necessity of considering the damping or particle radius and Stokes-Einstein relation.

  In the one-particle approximation, there is an information loss associated with integrating over the full N-particle distribution, $\rho(x,p,t)$.
Because less information is used to propagate the system, there is a start-up entropy production of $\H_\text{max}[\rho(x,p,t)|\rho_\alpha(r,t)]-\H[\rho(x,p,t)] \ge 0$ that is due to a loss of useful work that may be extracted from the system without this information.
Time-evolution followed by discarding information on particle correlations then transforms Eq.~\ref{eq:ent} into
\begin{equation}
 dS_c = \H_\text{max}[\rho(t+\epsilon)]-\H_\text{max}[\rho(t)] - dI_c^T\beta \ge 0
.
\end{equation}
For the Fokker-Planck equation of Brownian motion (\ref{eq:diffn}), the only information used to propagate the distribution is the density at each time-step, and the maximum entropy distribution is the momentum-free distribution with known average spatial density and average energy.\cite{wrobi90}  In the special case where the energy is a local function of the density, this leads to the well-known local equilibrium theory\cite{skjel08}.

  The momentum and non-diffusive conservation equations may also be derived from the full Langevin equation (\ref{eq:langevin}) in a manner similar to that shown in Ref.~\citenum{rleho10}.  Although more complex, these hydrodynamic systems are amenable to the analysis given above.  The entropy production will be a combination of heat production and information loss.  If the system moves between two steady-states, the entropy production will be given simply by the end-point information entropy difference plus the integral of the evolved heat divided by the temperature of the external heat reservoir.  These quantities are both linked to information loss because the microscopic details of the energy exchange have been replaced by less informative probability distributions over the states of both systems.

\subsection{ A Simple Application}

  For a numerical calculation, we may turn to the one-dimensional example of an optically trapped bead or an atomic-force microscope pulling experiment.  Assuming very fast relaxation of the momentum of the pulling coordinate over a potential energy surface, $U$, it is appropriate to use Eq.~\ref{eq:brown} with the average velocity set at zero.
In the case of an optical trap with center $x_0(t)$ and force $F(x,t) = -\kappa (x-x_0(t))$, we can employ a change of coordinates to $y(t)=x(t)-x_0(t)$ (with $\dot x_0(t) \equiv v(t)$) so that during each time-step, the position and trap center are updated to give
\begin{equation}
 \dot y \epsilon = -v(t)\epsilon -\frac{\kappa}{\gamma} y\epsilon + C_q^{1/2} \dW, \; \gamma^{-1} \equiv \beta C_q/2
\end{equation}
The constraint on $\dot x^2\epsilon/C_q$, determines the rate at which the bead is allowed to dissipate energy into the solution, while the energy constraint, $\beta$, acts as a driving force for net energy exchange.

  In the Brownian case, the system cannot distinguish between internal and external forces, so that Eq.~\ref{eq:dW} shows energy decrease as work from all applied (assumed internal) forces is dissipated into the surroundings as heat at each time-step.  Reversing our sign convention to treat the harmonic trap as external, the work done on the bath through the system is
\begin{align}
\label{eq:dWF}
 \avg{dW_F|y_i} &= \avg{F \dot x\epsilon|y_i} = -\avg{\kappa \bar y \dot x \epsilon|y_i} \\
 &= \frac{\kappa}{\gamma} (\kappa y^2 - 1/\beta)\epsilon,\notag \\
\intertext{Although we are not including this term in our analysis, if a potential had been present for the particle, at each step energy}
 \avg{dW_U|y_i} &= \avg{-\pd{U}{x} \dot x \epsilon|y_i} \\
\end{align}
would also have been converted into heat from the bath system's internal potential, $U$.%
\footnote{A pulling potential of mean force is also commonly used for $U$, showing that this term can be characterized as a coupling to an external thermostatic system.}

  The entropy increase of Eq.~\ref{eq:ent} (Eq.~\ref{eq:FT}) is formally a path average, and its evaluation requires specifying an initial state and a driving protocol.  As in Ref.~\citenum{etrep04}, we may choose to follow several velocity programs starting from a steady-state at constant pulling force.
\begin{equation}
\label{eq:smoF}
 \rho_\text{ss}(y) \propto e^{-\frac{\beta\kappa}{2}(y + \gamma v/\kappa)^2}
\end{equation}
The `housekeeping heat' dissipated by the bath's removal of the bead's momentum at each time-step leads to a steady-state dissipation of $\avg{dW_F}/dt = \gamma v^2$.
After a sufficiently long time, the instantaneous information entropy will reach its steady-state value, $\H[\beta,\kappa] = (1-\ln\tfrac{\beta\kappa}{2\pi})/2$.
Note that the entropy of the steady-state distribution is well-defined because it is invariant to the change of coordinates $x\to y$.
For constant pulling force and temperature, this expression says that over long time-periods the integral of the information entropy change will be zero, and the total entropy increase will be due completely to dissipated work.

\begin{figure}[htbp]
\includegraphics[width=3.33in]{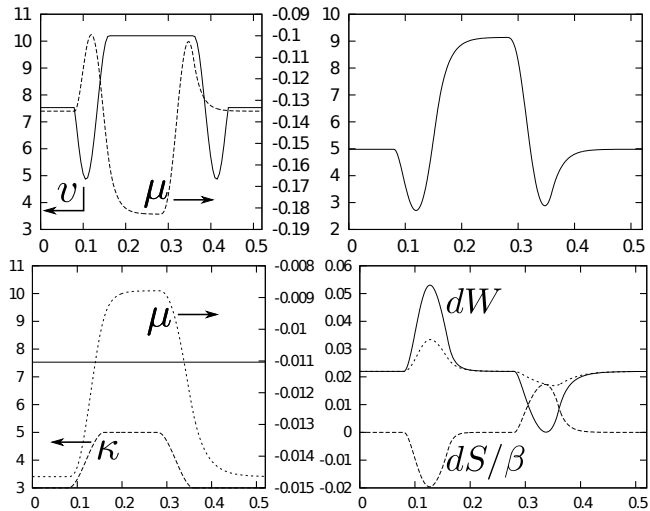}
\caption{Calculated entropy production during a transition between steady-states.  The left set of panels show the imposed velocity, $v$, solid line; force constant, $\kappa$, long dashed line on the left scale; and response of the mean, $\mu$, on the right {\em vs} time (s).  The right set of panels show the decomposition of the entropy increments (Eq.~\ref{eq:incr}, pN-$\mu$m) into heat (solid) and information gain/loss (long-dashed).  Whenever the information entropy decreases, an equal or larger amount of heat is produced so that the total (Eq.~\ref{eq:totSF}, short-dashed) is always positive.  For the upper two sets of panels, the force constant was held constant at $\kappa=4.9$ pN/$\mu$m so no change in information entropy occurs.  For the lower two, the distribution is compressed, then broadened by changing $\kappa$ between 3 and 5 using a cubic interpolation lasting 80 ms.}
\label{fig:drive}
\end{figure}

  During intermediate time-periods, however, the entropy increase will be a combination of changes in the position distribution plus the dissipated work.
\begin{equation}
\label{eq:incr}
 dS_i = d\H_i - \beta\avg{dE}_i = \avg{-\ln\frac{\Pri{i+1}{y_{i+1}}}{\Pri{i}{y_i}} + \beta dW_F(y_{i+1};y_i)}_i
.
\end{equation}
Because the dynamics is Markovian, the average dissipated work can be easily calculated from the distribution at each time-step and the total of each type of work will be a sum of one-step stochastic integrals (Eq.~\ref{eq:dWF}).  Assuming the distribution is Gaussian at a starting time and Fourier-transforming Eq.~\ref{eq:smoF} gives a Gaussian distribution at all future times for all driving protocols, whose mean ($\mu$) and variance ($w$) are solutions of first-order ordinary differential equations.
\begin{align*}
\frac{d\mu(t)}{dt} &= -(v+\frac{\kappa\mu}{\gamma}),&
\frac{dw(t)}{dt} &= 2D(1-\beta\kappa w) \\
\intertext{The information entropy and work follow}
\frac{d\H}{dt} &= D(\frac{1}{w}-\beta\kappa),&
\frac{dW_F}{dt} &= \frac{\kappa}{\gamma}(\kappa(\mu^2+w)-\tfrac{1}{\beta})
.
\end{align*}
Using $x+x^{-1} \ge 2$, we can easily verify
\begin{equation}
\label{eq:totSF}
\avg{dS_i} = D\beta^2\kappa^2\mu^2 + D\beta\kappa\left[\beta\kappa w + (\beta\kappa w)^{-1} - 2 \right] \ge 0
.
\end{equation}
Using numerical integration, we have plotted (Fig.~\ref{fig:drive}) the rate of entropy production vs. time for two hypothetical driving protocols.  The units in the figure are the same as in Ref.~\citenum{etrep04}, and their third protocol (followed by its time-reversal starting at 0.28 s) has been used for the upper two sets of panels.  Because the variance of the distribution only responds to changes in diffusion constant, temperature or driving force, we have varied $\kappa$ in the second set of calculations.  The information entropy rate gain goes to zero and the heat production becomes constant at the onset of the eventual steady-state.  As shown by the heat production during compression from $\kappa=3$ to 5 pN/$\mu$m, excess heat production is required whenever the information entropy decreases.  When the distribution expands, heat production decreases as the mean begins to lag behind the trap center and the distribution expands.  This is counter-balanced by an increase in information entropy, leading to net dissipation.

  We note that a large amount of additional complexity can be added to this model by adding information about the variables here treated as `external' to the description of the dynamics.  If local variations in the fluid velocity or temperature were included, then the dynamics would have to specify the equations of motion for these fields.  The final entropy increase may then be more or less than this result because these degrees of freedom may be responsible for additional heat production, but more information on the fluid state has been included, leading to decreased information loss.

\section{ Conclusions}

  In this paper, we have given a generalization of the theory for
driven, irreversible processes.  A set of transition probabilities defines
the evolution equation for the system of interest.  A special simplification is the case of Langevin and
Brownian motion, which can be recovered as limits of a constrained action
integral approach.  Non-anticipating stochastic trajectories for classical particle and field motion can
be cast in this form.  The action functional interpretation gives a physical
method for defining conserved quantities and the energy cost associated with
transfers of these quantities from an external environment or experimental apparatus.

  Deterministic dynamics is recovered from the Langevin equation when
the deviation of the action functional is strongly constrained to zero
(Sec.~\ref{sec:dyn}).  In this limit, the external forces which appeared as
statistical in the stochastic approach become mechanical.  Because both limits
appear in this derivation, the fluctuation-dissipation theorems derived
as Gibbs relations from Eq.~\ref{eq:one} are applicable in the case of both
thermal and mechanical driving forces.  These equations are completely general
in the sense that they apply not only arbitrarily far from equilibrium, but
also during transient processes which do not possess a steady state.


  A particularly useful aspect of this approach is that it directly connects
multiple length and time-scales.  The formulation of the equations has been
in terms of particle motion, but coarse-grained relations are easy to define
as appropriate ensemble averages over these motions.  Examples of such
averages include centers of mass for polymer units or average density and
velocity fields.  The coarse equations of motion
will then lead to polymer coarse-graining models\cite{rakke00} or non-local hydrodynamic
models.\cite{wgran87,rkubo66,rleho10}
For the time-evolution of average quantities, we expect the thermodynamic
limit argument\cite{ejayn57a} to apply when the number of averaged degrees of
freedom is large so that the path realized by the system
under a given set of constraints will fall arbitrarily close to the maximum
entropy solution an overwhelming majority of the time.
The present work is therefore a suitable foundation for the theory and
analysis of nonequilibrium molecular dynamics.

  Applications to simplified, standard examples such as circuit theory are
easily accomplished.  The Joule heating of a resistor, for example, can be seen
from Eq.~\ref{eq:pdiss} as fundamentally arising from the difference between
the velocity added to each ion individually {\em vs.} the usable energy
in the average ion velocity.  Because the energy added to the system in
driving the ions is not expressible in terms of the average velocity alone,
spreading in the distribution of ion velocities becomes heat.  The same
remarks follow for driven convective transport, where a spreading in the
distribution of forward fluid momentum leads to increases in the local
temperature (Eq.~\ref{eq:temp}).


  Connections of this theory to the formal structure of maximum entropy
thermodynamics and Bayesian inference have been elaborated upon in Ref.\citenum{droge11a}
These connections allow the definition of
thermodynamic cycles expressing differences between driving protocols using
the same free energy techniques commonly employed in the equilibrium theory.
Some examples have already appeared in the literature for path
re-weighting\cite{dzuck99,dminh09}.  It is expected that expression in terms of
thermodynamic cycles will greatly simplify the derivation and interpretation
of these studies.

  We have identified a new generalization of the second law for irreversible
processes.
A traditional analysis shows that the total entropy increase
(Eq.~\ref{eq:ent}) is dependent on the details of system dynamics and exchange of conserved quantities with an external system.  Connecting this with the fluctuation theorem (Eq.~\ref{eq:FT}) gives a microscopic form for the second law of thermodynamics.  The physical device of tracking work performed on individual particles as well as external reversible work sources allows us to track the flow of each type of work (and heat) through the system.  Because these changes come directly from the forces on each degree of freedom, this analysis does not depend arbitrary decompositions of energy functions or definitions of steady-states.

  From an informational perspective, entropy increase comes about from
discarding information and/or from the information loss associated with
coupling to external reservoirs.  This is distinguished from the entropy
production functional of local equilibrium theory in that the entropy
functionals developed here include long-range correlations and are not
necessarily extensive.\cite{ejayn92}
It is a simple matter to define more complicated baths, for example affecting only the average temperature in a given area for imposing thermal gradients.
It should be noted that the analysis in Sec.~\ref{sec:thermo} showed that increasing the number of controllable variables decreases the number of degrees of freedom associated with heat production.
Molecular insertion and deletion operations will aid in generalizing
this approach to include imposed chemical potential (insertion force) as well as particle flux boundary conditions, but have not been considered here.

\section*{Acknowledgment}
  This work was supported, in part, by Sandia's LDRD program.
Sandia National Laboratories is a multi-program
laboratory operated by Sandia Corporation, a wholly owned subsidiary
of Lockheed Martin Corporation, for the U.S. Department of Energy's
National Nuclear Security Administration under contract DE-AC04-94AL85000.

\appendix
\section{ Analytical Calculation of Stochastic Integrals}

  Despite the wealth of literature on the Langevin and Wiener processes, the
procedure for calculating expectations of time-integrals given in standard
references such as Gardner\cite{cgard04} and Risken\cite{hrisk96} remains complicated.  Because Stratonovich integrals appear prominently in the present paper, often usurping the role of thermodynamic potentials, we present here two alternative methods.  Both rely on replacing expressions to be evaluated at the midpoint of a time-step with the first-order expansion, $f(\bar x) \approx f(x) + \pd{f(x)}{x}\dot x\epsilon/2$.

  Using Eq.~\ref{eq:dI} with $Y_0=\bar{\dot x}$ to find the energy change, we expand the average velocity about the mid-point,
\begin{align}
 \bar{\dot x} &= \frac{\dot x_{i+1}+\dot x_i}{2} = \dot x_i + \epsilon/2 M^{-1} \dot p \\
 &= \dot x_i + \tfrac{1}{2} M^{-1} \left(F\epsilon - \tfrac{\epsilon}{2}C\tilde Y\tilde\beta
     - \tfrac{\epsilon}{2}C\bar{\dot x}\beta_0 + C^{1/2}\dW \right) \\
 &= (I+\tfrac{\epsilon\beta_0}{4}M^{-1}C)^{-1}\left[ \dot x_i
   + \tfrac{\epsilon}{2}M^{-1}(F-C\tilde Y\tilde\beta/2) + \tfrac{1}{2}M^{-1}C^{1/2}\dW \right] \\
 &= (I-\tfrac{\epsilon\beta_0}{4}M^{-1}C)\dot x_i
   + \tfrac{\epsilon}{2}M^{-1}(F-C\tilde Y\tilde\beta/2) + \tfrac{1}{2}M^{-1}C^{1/2}\dW + O(\epsilon^{3/2}) \\
 &= \dot x_i + \tfrac{\epsilon}{2} M^{-1}\dot p' + O(\epsilon^{3/2})
,
\end{align}
where $\dot p'$ is computed using only quantities at the time-step $i$.  Multiplying this with $-\dA{t}$ from Eq.~\ref{eq:langevin}, we get
\begin{equation}
 dI_0 = \bar{\dot x}(\dot p-F)\epsilon = \tfrac{1}{2}\dW^TC^{1/2}M^{-1}C^{1/2}\dW + \dW^T C^{1/2}\dot x_i
 - \tfrac{\epsilon}{2}\beta^T{Y'}^TC\dot x_i + O(\epsilon^{3/2}).
\end{equation}
Since $M^{-1/2}C^{1/2}\dW$ is normally distributed with mean zero and variance-covariance matrix $M^{-1/2}CM^{-1/2}\epsilon$, $dI_0$ has a noncentral $\chi^2$ distribution with expectation
\begin{equation}
 \tfrac{\epsilon}{2}\left[\operatorname{Tr}(M^{-1} C) - \beta^T {Y'}^T C \dot x_i\right]
\end{equation}
For a single constraint, $Y'=\dot x$, we find a definition of the kinetic temperature, Eq.~\ref{eq:temp}.  If, in addition, we include a constant pulling force, $\beta_1=-\lambda$, we find
\begin{equation}
\label{eq:langen}
 \avg{dI_0} = \tfrac{1}{2}\left[\operatorname{Tr}(M^{-1}C)
  -\beta_0(\lambda/\beta_0\openone-\dot x_i)^T C (\lambda/\beta_0\openone-\dot x_i)
  +\lambda(\lambda/\beta_0\openone-\dot x_i)^T C \openone \right]
.
\end{equation}
What emerges is a kinetic temperature with respect to the terminal velocity, $\lambda/\beta_0$, as well as a heating term.

  This method can also be used to prove Eq.~\ref{eq:dWF}, starting from the expansion
\begin{align}
\bar y &= y_i + \epsilon\dot y/2 \\ 
 &= y_i - \tfrac{1}{2}\left(v(t)\epsilon+\tfrac{\kappa}{\gamma}\bar y\epsilon-C^{1/2}\dW\right) \\
 &= (1+\tfrac{\kappa\epsilon}{2\gamma})^{-1}(y_i-\tfrac{v\epsilon}{2}+\tfrac{1}{2}C^{1/2}\dW) \\
 &= (1-\tfrac{\kappa\epsilon}{2\gamma})y_i - \tfrac{v\epsilon}{2} + \tfrac{1}{2}C^{1/2}\dW + O(\epsilon^{3/2})
.
\end{align}

  As discussed in the text, these integrals should also result from differentiating a partition function (Eq.~\ref{eq:F}).  We present an extended derivation of the main results of this approach here.  Both Langevin and Brownian equations can be derived as appropriate limits of the constraints
\begin{align}
\eta(t)/\epsilon =
\begin{bmatrix}
 \frac{\partial \mathcal A}{\partial q} \\
 \frac{\partial \mathcal A}{\partial v} \end{bmatrix}^T
\begin{bmatrix}
 G_p & \\
  & G_q \end{bmatrix} \left(
\begin{bmatrix}
 \frac{\partial \mathcal A}{\partial q} \\
 \frac{\partial \mathcal A}{\partial v} \end{bmatrix}
- \begin{bmatrix}
  \dot q\beta/2 + g(\bar q)\lambda/2 \\
  \dot v\beta/2 + h(\bar q)\lambda/2 \end{bmatrix} \right)
,
\end{align}
where $\lambda g(q)$ and $\lambda h(q)$ introduce external forces.  Next, we make the one-half step substitutions,
\begin{equation}
\begin{bmatrix}
 \frac{\partial \mathcal A}{\partial q} \\
 \frac{\partial \mathcal A}{\partial v} \end{bmatrix} \to
\begin{bmatrix}
 M(f(q)+F(q)\dot q\epsilon/2 - \dot v) \\
 M(\dot q-v-\dot v\epsilon/2) \end{bmatrix},\quad
g(\bar q) \to g(q) + G(q)\dot q\epsilon/2
,
\end{equation}
using the force-per-mass, $f$, the appropriate derivative matrices $\{F,G,H\}_{IJ} = \partial \{f,g,h\}_I / \partial q_J$, and defining $J_q\equiv M G_q M\equiv C_q^{-1}/2$, $J_p\equiv M G_p M\equiv M C^{-1} M/2$.
Factoring $\eta$, gives a normal distribution for $[\dot v,\dot q]^T\epsilon$ with penalty matrix (inverse of the variance-covariance matrix, keeping terms below $O(\epsilon)$)
\begin{equation}
P = \begin{bmatrix}
J_p+\epsilon M\beta/4 & \epsilon (M G\lambda/4-F^T J_p-J_q)/2 \\
\epsilon (G^TM\lambda/4-J_p F-J_q)/2 & J_q - \epsilon (MH\lambda+MF\beta)/4 \end{bmatrix}/\epsilon
\end{equation}
and mean (to first order in $\epsilon$)
\begin{equation}
\label{eq:motion}
\avg{\begin{bmatrix}\dot v\epsilon \\ \dot q\epsilon \end{bmatrix}}
 = \begin{bmatrix}
  f - M^{-1}C (\tfrac{\beta}{2}v + \tfrac{\lambda}{2}g) \\
  v + C_q M(\tfrac{\beta}{2} f + \tfrac{\lambda}{2} h) \end{bmatrix} \epsilon
.
\end{equation}
These expressions are in accord with Eqns.~\ref{eq:langevin} and~\ref{eq:brown}.

  The residual terms contribute to form the transition free energy functional (again to order $\epsilon$),
\begin{align}
 \mathcal F = \tfrac{1}{2}\big[ \ln \tfrac{|2\epsilon P|}{(2\pi)^d}
 &- (\beta v+\lambda g)^T C (\beta v + \lambda g)\epsilon/4 \\
 &- (\beta Mf+\lambda Mh)^T C_q (\beta Mf+\lambda Mh)\epsilon/4 \big]
.
\end{align}

  Note that the Fokker-Planck equation can be used to prove that the Boltzmann distribution is stationary under either the Langevin ($C_q\to 0$) or Brownian ($C\to 0$) limits, but not both.
For the Langevin limit, it can be checked that the derivative of this equation with respect to $\beta/2$ gives Eq.~\ref{eq:langen}.  For the Brownian limit, we find again Eq.~\ref{eq:dWF}.  These rely on the following expansion for the derivative of the log-determinant term
\begin{align*}
 \pd{\ln |\epsilon P|}{\alpha} &= \operatorname{Tr}\left(
    (\epsilon P)^{-1}\pd{(\epsilon P)}{\alpha}\right) \\
\epsilon P &= P_0+\epsilon P_1 + O(\epsilon^2),\; P_0 \equiv
\begin{bmatrix} J_p &  \\ & J_q \end{bmatrix} \\
(\epsilon P)^{-1} &= P_0^{-1} - \epsilon P_0^{-1}P_1P_0^{-1} + O(\epsilon^2)
.
\end{align*}
Since $\pd{(\epsilon P)}{\alpha}$ should contain a prefactor of $\epsilon$, the second term is usually unimportant, so that
\begin{equation}
 \pd{\ln |\epsilon P|}{\alpha} = \operatorname{Tr}\left(
2 \begin{bmatrix} M^{-1}CM^{-1} &  \\ & C_q \end{bmatrix}
\pd{(\epsilon P)}{\alpha}
\right).
\end{equation}

\bibliographystyle{unsrt}

\end{document}